# New techniques in Monte Carlo simulation: experience with a prototype of generic programming application to Geant4 physics processes

Maria Grazia PIA[1*], Mauro AUGELLI[2], Marcia BEGALLI[3], Lina QUINTIERI[4], Paolo SARACCO[1], Manju SUDHAKAR[1], Georg WEIDENSPOINTNER[5], Andreas ZOGLAUER[6]

[1] *INFN Sezione di Genova, 16146 Genova, Italy*
[2] *CNES, 31401 Toulouse, France*
[3] *State University Rio de Janeiro, 20550-013 Rio de Janeiro, Brazil*
[4] *INFN Laboratori Nazionali di Frascati, 00044 Frascati, Italy*
[5] *MPE and MPI Halbleiterlabor, 81739 München, Germany*
[6] *University of California at Berkeley, 94720 Berkeley, CA, USA*

An investigation is in progress to evaluate extensively and quantitatively the possible benefits and drawbacks of new programming paradigms in a Monte Carlo simulation environment, namely in the domain of physics modeling. The prototype design and extensive benchmarks, including a variety of rigorous quantitative metrics, are presented. The results of this research project allow the evaluation of new software techniques for their possible adoption in Monte Carlo simulation on objective, quantitative ground.

## I. Introduction

The Geant4[1)2)] toolkit provides advanced functionality for all the domains typical of detector simulation: geometry and material modeling, description of particle properties, physics processes, tracking, event and run management, user interface and visualization. Geant4 is nowadays a mature Monte Carlo system and is used in many, multi-disciplinary experimental applications; its rich collection of physics processes and models, extending over a wide energy range, has played a key role in satisfying the needs of a large variety of experimental developments.

New technological developments in software and computing hardware have also occurred since the RD44[3)] phase, which defined Geant4 design. New software techniques are available nowadays, which were not yet established at the time when Geant4 was designed.

The project in progress[4)] described here studies the implications, possible advantages and drawbacks of using new techniques in simulation design; for this purpose a prototype investigates the adoption of generic programming techniques in Geant4 electromagnetic physics domain.

## II. Generic programming techniques in Monte Carlo simulation

New, powerful programming paradigms have emerged in recent years, like generic programming[5)] and template metaprogramming[6)]. In C++[7)] these techniques are enabled by the flexibility of templates[8)], the C++ type parameterization mechanism. In C++ the template mechanism provides naturally a rich facility for the application of these programming paradigms.

These techniques have not been exploited at large scale in any major Monte Carlo system yet. The use of templates was not mature for generalized adoption in a large scale, multi-platform simulation system at the time of the RD44 phase of Geant4 design: the evolution towards the C++ standard[9)] was still in progress, and the limited support of templates in C++ compilers in the mid 90's discouraged their exploitation as major players in Geant4 architectural design.

Various features of generic programming identify it as a worthy candidate for physics simulation design.

One of the main advantages of template code is its capability of accommodating multiple options: this is indeed a requirement in many Monte Carlo systems, where multiple physics modeling options are provided. A small amount of template code can support many implementation variants in a consistent, extensible and maintainable fashion.

Generic programming focuses on finding commonality among similar implementations of the same algorithm, then providing suitable abstractions so that a single, generic algorithm can cover many concrete implementations. A design based on this technique would naturally overcome issues of "duplicated" or "competing" functionality in different physics packages of the same simulation system, which is often the result of the evolution of the original code into multiple specialized implementations. The concepts that emerge from the process of finding commonality across multiple implementations contribute to better understanding the problem domain: the importance of this side benefit in physics modeling should not be underestimated.

---

*Corresponding Author, E-mail:zzz@abc.co.jp

Customization and extensibility through the provision of user-specific (or experiment-specific) functionality in the simulation are also facilitated.

A side product of the adoption of generic programming techniques in Monte Carlo simulation design is the improved transparency of physics models: the technology intrinsically achieves their exposure at a fine-grained level. This feature greatly facilitates the validation of the code at microscopic level and the flexible configuration of physics processes in multiple combinations; it also contributes to expose any epistemic[10] uncertainties affecting low-level building blocks of physics models, which would propagate their effects into experimental physics observables.

Also the usage of physics modeling options of the toolkit in experimental applications is facilitated: in fact, generic programming allows the user to write more expressive code, that more closely corresponds to the mental model of the problem domain.

An advantage of generic programming over conventional object oriented programming is the potential for performance improvement. Physics modeling would profit from a paradigm shift, which would exploit static polymorphism to provide a variety of modeling options at the place of dynamic polymorphism, as it is currently the case in an object oriented system like Geant4: the former binds the choice of a physics option at compile time rather than runtime, thus resulting in intrinsically faster programs. Design techniques intrinsically capable of performance gains are relevant to computationally intensive simulation domains, like calorimetry and microdosimetry: the development of high energy electromagnetic shower and detailed particle treatment down to very low energy according to discrete transport schemes, which characterize these applications, are especially demanding towards the computational performance of physics simulation software. In general, the large scale simulation productions required by HEP experiments would also profit from opportunities for improved physics performance.

Generic programming exhibits some drawbacks along with benefits; they affect both the development of the software and its use.

Many compilers historically have poor support for templates, thus the use of templates can make code somewhat less portable: this drawback can be significant for general purpose Monte Carlo systems, which should run on a variety of platforms. Support may also be poor when attempting to use templates across shared library boundaries. Most modern compilers however now have fairly robust template support, and the new C++ standard is expected to further address these issues.

From the perspective of software development, compilers often produce confusing, long and unhelpful error messages when errors are detected in code that uses templates. The difficulty of debugging the code can make code based on the generic programming paradigm difficult and time consuming to develop.

From the user's perspective, the major drawback of using templates is code bloat. The use of templates requires the compiler to generate a separate instance of the templated class or function for every permutation of type parameters used with it. Therefore indiscriminate use of templates can lead to code bloat, resulting in a very large executable. However, clever use of template specialization can limit code bloat in some cases.

### III. Investigation of a prototype

Generic programming looks a promising candidate for exploitation in Monte Carlo physics modeling; nevertheless, the drawbacks exhibited by this technique are source of concern, since they represent severe risks for large scale, widely used, multi-platform Monte Carlo systems. The risks are even more serious in situations where a Monte Carlo system is used in production mode by experiments at data-taking time, as it is the case of the experiments at the Large Hadron Collider (LHC) currently operating at CERN.

Preliminary investigation of the applicability of generic programming techniques in Monte Carlo simulation[11] was performed by one of the authors of this paper in a small physics sub-domain of Geant4; although this first experience was successful at that limited scale, the exploitation of this technique is not established yet for application to a large scale physics simulation domain.

This yet unclear picture of the potential benefits and drawbacks of generic programming application to Monte Carlo physics modeling has been addressed with a scientific attitude: a pilot project[12] studies the implications, possible advantages and disadvantages of this software design technique in a concrete prototype concerning a relatively large and complex physics simulation domain, with the purpose of collecting significant elements to evaluate the suitability of generic programming techniques to physics simulation design. The project is rigorously managed to acquire and document quantitative metrics; its results would allow Monte Carlo simulation developers and users to evaluate the benefits and drawbacks of the new paradigm on objective ground and would enable the choice of an optimal programming paradigm for that physics simulation domain in scientific terms. This process conforms to the principles of academic freedom in scientific research[13], which are at the basis of the development of many major Monte Carlo systems.

The first development cycle of the pilot project in progress is focused on electromagnetic physics, namely on photon interactions. The research on new programming paradigms concerns the physics processes and models of photon interactions as currently available in Geant4; the effort at this stage is focused on their software design aspects, while further improvements or extensions of physics functionality are not within its scope, at least at this early stage of the research process.

The software development process is based on an iterative and incremental model, which adopts the Unified Process[14] as a software process framework. The adopted process has been tailored to the specific characteristics of the project and is continuously improved, based on the experience gained in the course of the development.

The main concepts driving the design are: flexible configuration of processes at granular level, performance optimization, transparency of physics and facilitation of the software test process (both verification and validation).

At the present stage a policy-based class design[15] has been adopted; the investigations currently in progress evaluate whether it meets these requirements, estimate quantitatively the benefits and drawbacks of this technique with respect to the application environment, and assess the advantages – or disadvantages – it would offer with respect to conventional object oriented design methods and to the current Geant4 software implementations.

The prototype design follows a minimalist approach: a generic process acts as a host class, which is deprived of any intrinsic physics functionality; physics behaviour is acquired through policy classes. A physics process is independent not only from the physics models that determine the cross-section and final state generation, but also from their types.

The main characteristics of this approach are:
- a policy defines a class or a class template interface
- policy host classes are parameterized classes
- policies are not required to inherit from a base class (but they may, if appropriate)

In the initial stage two basic policies have been identified and defined; they are respectively responsible for cross section and final state generation. An example of the policy-based design is shown in **Figure 1** in UML[16] (Unified Modeling Language); it is stressed that this design is still preliminary and subject to further evolution.

Implementations conform to the policies through "duck typing". Current Geant4 models of photon interactions implemented in Geant4 standard[17] and low energy[18,19] packages (the latter based on data libraries and on physics algorithms originally developed for Penelope[20]) have been reimplemented according to this design scheme.

The code is bound at compilation time: since no virtual methods are needed, faster execution is expected. Preliminary performance measurements in a few simple test cases concerning photon interactions indicate a gain on the order of 30% with respect to equivalent physics implementations in the current Geant4 design scheme. Further tests are needed, however, to acquire a more extensive and in-depth understanding of the computational performance and its potential drawbacks in terms of code bloat. It should also be stressed that no effort has been invested yet into optimizing the new design prototype, nor the code implementation.

The prototype design scheme greatly facilitates the test of the software: physics functionality is associated with low level objects like policy classes, which can be verified and validated independently. This agility represents an improvement over some heavy-weight design schemes, where a full-scale Geant4-based application is necessary to study even low-level physics entities of the simulation, like atomic cross sections or features of the final state models. For instance, the same tests[21] comparing cross section implementations to NIST physical reference data[22] profit from approximately a factor 100 reduction in the number of lines of code; the production effort scales from order of weeks in a computing farm, with a dedicated production manager to order of minutes of human time on a laptop computer.

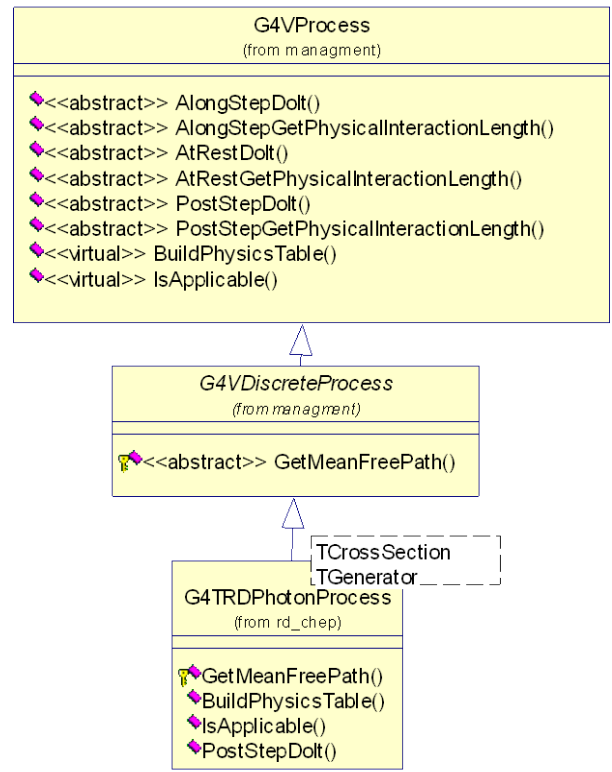

**Fig. 1** UML class diagram illustrating the main features of the policy-based prototype design.

Metrics concerning the development effort have been collected as well, and compared to previous experience of implementing Geant4 electromagnetic processes according to conventional object oriented methods. In both cases the implementation was performed by novice developers, who were at their first experience of code development for a Monte Carlo simulation kernel. Preliminary results indicate that the development in a policy-based class design is an order of magnitude faster than in a conventional object-oriented environment; this result can be explained by the great simplicity of the design. Code reviews are also faster for the same reason; preliminary appraisals indicate approximately three times faster reviews with respect to reviewing the original Geant4 code. It is worthwhile to remind the reader that code reviews are key practice for the quality of the software itself[23]. Metrics concerning the design effort have been collected as well; preliminary indications suggest slightly larger investment in design adopting generic programming with respect to conventional object oriented design, but more extensive tests are needed to acquire reliable statistics in this respect, and to evaluate possible bias in the results due to different previous experience with the two techniques, and to external factors affecting the efficiency and the mental serenity of the designer in the two cases.

It is worthwhile to recall that, since dynamic and static polymorphism coexist in C++, the adoption of generic programming techniques would not force the developers and users of the new code to replace object oriented methods entirely: a clever design can exploit generic and object oriented programming techniques in the same software environment, as most appropriate to the characteristics of the problem domain.

### III. Conclusion

This research study has been performed over a wide physics domain – photon interaction processes, which involve a variety of modeling approaches and options. It aims at providing adequately articulate feedback about the technical implications of the new programming paradigms over several simulation features: flexibility of configuration, computational performance and memory usage, maintainability, facilitation of verification and validation, simplicity of use in experimental applications.

The first development cycle is focused on the use of policy-based class design in the domain of photon interactions. A prototype design and implementation has been followed by benchmarks including a variety of rigorous quantitative metrics.

The first results indicate that this technique is suitable to support the design of the discrete simulation sector in an efficient, transparent and easily customizable way. The lightweight and easily manageable design achievable with such techniques would greatly facilitate further evolution to accommodate a variety of functionality. Further developments and benchmarks are planned to better characterize the implications of new design paradigms in Monte Carlo physics domains.

The metrics collected in the course of this research project allow the evaluation of new software techniques for their possible adoption in Monte Carlo simulation on objective, quantitative ground.

### Acknowledgment

The authors express their gratitude to CERN for support to the research described in this paper.

The authors thank Sergio Bertolucci, Thomas Evans, Elisabetta Gargioni, Simone Giani, Vladimir Grichine and Andreas Pfeiffer for valuable discussions.